\documentclass{elsart}
\usepackage[]{epsfig}
\usepackage{pstricks}
\begin{document}
 
\begin{frontmatter}

	\title{Dynamically self-regular quantum harmonic black holes}
	\author{Euro Spallucci\thanksref{infn}}
\thanks[infn]{e-mail address: spallucci@ts.infn.it }
\address{Dipartimento di Fisica Teorica, Universit\`a di Trieste
and INFN, Sezione di Trieste, Italy}
 
\author{Anais Smailagic\thanksref{infn2}}
\thanks[infn2]{e-mail address: anais@ts.infn.it }
\address{Dipartimento di Fisica Teorica, Universit\`a di Trieste
and INFN, Sezione di Trieste, Italy}
        
	\begin{abstract}
	The recently proposed UV self-complete quantum gravity program is a new and very interesting  way to envision
        Planckian/trans-Planckian physics. in this new framework, high energy scattering is dominated by the creation of
        micro black holes, and it is experimentally impossible to probe distances shorter than the horizon radius.
        In this letter we present a model which realizes this idea through the creation of  \emph{self-regular} 
        quantum black holes admitting a minimal size  extremal 
        configuration. Their radius provides a \emph{dynamically generated minimal length} acting 
        as a universal short-distance cutoff. We   propose a quantization scheme for this new kind of microscopic 
        objects based on a Bohr-like approach, which does not require a detailed knowledge of quantum gravity. 
        The resulting black hole quantum picture resembles the energy spectrum 
        of a quantum harmonic oscillator.  The mass of the extremal configuration plays the role of zero-point energy. 
        Large quantum number re-establish the classical black hole description. Finally,
        we also formulate a ``quantum hoop conjecture'' which is satisfied by all the mass eigenstates and sustains
        the existence of quantum black holes sourced by Gaussian matter distributions. 
	\end{abstract}
	\end{frontmatter}
	
\section{Introduction}

The idea that non-perturbative quantum gravity can ``cure''  ultraviolet divergences, including its own,
 dates back to the seventies \cite{Delbourgo:1969ha,Salam:1970bd,Isham:1970aw}, the idea was  further developed by 
\cite{SriRam:1979fr} and very recently embodied in the
 so called ``UV self-complete quantum gravity'' by Dvali and collaborators \cite{Dvali:2010bf,Dvali:2010jz}.
 The novelty of this
approach consists in the assumption that Planckian energy scattering will be dominated by the production of
micro black holes (BHs). So far, the paradigm of
modern high-energy physics is that the energy of an accelerated particle allows to probe
shorter and shorter distances without any lower bound. The present LHC peak energy, $14\, TeV$, set the experimental
limit up to $10^{-17} \, cm $. Hypothetically, an ultra-Planckian particle accelerator would be even
able to probe  distances below $10^{-33} cm$. Although  there is no chance to build such a machine in a foreseeable
future, the theoretical argument remains valid.\\
Nevertheless, if one considers the collision of two elementary particles with high enough center of mass energy and small  
impact parameter,  a huge energy concentration would be reached requiring, according to UV self-complete quantum gravity
hypothesis, a proper account of non-perturbative gravitational effects. Such a situation is expected to lead to
the creation of a micro BH, as a realization of the ``quantum hoop conjecture'' (QHC).  QHC  extends the 
classical statement that a
macroscopic object of arbitrary shape, of mass $M$,  passing through a ring of radius $R=2MG$, will necessarily collapse 
into a BH \cite{thorne}. In the quantum case \footnote{So far, there is no  unique formulation
 of QHC. For an alternative definition see \cite{Casadio:2013uga}.}, 
the macroscopic object is replaced by the  target-projectile pair and the condition for a BH creation
is $2\sqrt{s}\, G \le b$, where $\sqrt{s}$ is the total center of mass energy of the colliding system, $G$ the gravitational
coupling constant and $b$ is the impact parameter. Thus, whenever the effective Schwarzschild radius is lower or equal to the
impact parameter, the BH production channel opens up. If Planckian scattering regime  is BH creation dominated
\cite{Rizzo:2002kb,Chamblin:2002ad,Mocioiu:2003gi,Lonnblad:2005ah,Rizzo:2006di,Najafabadi:2008xv,Chamblin:2008ec,Erkoca:2009kg,Kiritsis:2011qv},
the idea  higher-energy/shorter-distance needs a  substantial revision 
\cite{Spallucci:2011rn,Mureika:2011hg,Spallucci:2012xi,Nicolini:2011fy,Aurilia:2013mca,Nicolini:2012fy,Nicolini:2013ega}. 
In other words,  increasing $\sqrt{s}$ instead of reaching lower and lower wavelengths, 
stops at the threshold of BH creation. Any
further energy increase leads to growing BHs, thus shielding distances below their horizon  from experimental reach. \\
This idea has been recently incorporated in the framework of  large extra-dimension models, where quantum gravity effects
are expected to be dominant around $10-100\, TeV$. In this scenario ''$TeV$ BHs``  production is 
 slightly above the LHC energy and hopefully reachable by the next generation particle colliders. 
Far beyond this energy scale, contrary
to the standard expectation, gravitational dynamics becomes classical again (''large BHs``), thus ultra-Planckian regime 
is unexpectedly dominated by \emph{classical} field configurations, 
i.e. ''~classicalons~`` \cite{Dvali:2010jz,Dvali:2011th,Dvali:2012mx,Dvali:2014ila}.\\
 If  minimal size BHs are to be produced, their  length sale will serve a natural short-distance cut-off, i.e. 
a \emph{minimal length }.  The existence of a minimal length, $l_0$, in the space-time fabric was also implied 
by different approaches to quantum gravity including string theory, loop quantum gravity, non-commutative geometry, etc. 
\cite{book1}.
From a conventional point of view, $l_0$ is identified with the Planck length $l_P=\sqrt{G}$, but as described above
could be lowered near $TeV$ scale. 
Therefore, it can be expected that $TeV$ BHs should be sensible to the presence of $l_0$.\\
 In a series of papers we have
given BH solutions naturally incorporating $l_0(=\sqrt{\theta})$, where $\theta$ is a  parameter measuring the amount of
coordinate non-commutativity at short distance. 
In other papers, $l_0$ was engraved in the space-time fabric through
a $\ast$-product embedded into the very definition of the  metric tensor  $g_{\mu\nu}$, in terms of 
the vierbein field $e^a_\mu$:

\begin{equation}
 \hat{g}_{\mu\nu}\equiv \eta_{ab} \, e^a_\mu \ast e^b_\nu
\end{equation}

The latter approach faces  the basic difficulty that any attempt to solve the Einstein equations requires 
a truncated perturbative expansion in $l_0$, leading to 
an effective field theory with derivative couplings of arbitrary order. 
The resulting Feynman expansion still contains planar graphs which are divergent one by one,
in spite of the presence of $l_0$ \cite{Moffat:2000fv,Moffat:2000gr}. 
This is the consequence of the (truncated) perturbative treatment which changes the original meaning of $l_0$ 
from a natural UV cut-off into the (dimensional) strength of non-renormalizable derivative interactions. 
The difficulty with the perturbative treatment of the $\ast$-product can be summarized as follows:
in spite of the presence of $l_0$ in the theory, some of the resulting Feynman diagrams remain divergent.
To have a genuine non-perturbative approach we argued that 
the effects of $l_0$ can be implemented correctly   in General Relativity
by  keeping the standard form of the Einstein tensor in
the l.h.s. of the field equations and introduce an energy-momentum tensor with a modified source 
\cite{Nicolini:2005vd,Rizzo:2006zb,Ansoldi:2006vg,Nicolini:2008aj,Spallucci:2009zz,Smailagic:2010nv}.
The resulting solution for \emph{neutral, non-rotating,} BH  exhibits:
\begin{itemize}
 \item ''\emph{regularity}``, i.e. absence of curvature singularities;
 \item extremal configuration corresponding to a minimal size near $l_0$.
\end{itemize}
  
Regularity is an immediate consequence of the presence of $l_0$ in the space-time geometry, while the existence of a minimal
mass, extremal configuration, is a surprising property, at least  from the point of view of the BH textbook solutions.  \\

In the first part of this letter we present the regular Schwarzschild solution that exhibits
extremal configuration with radius $r_0=l_0$. This is what one expects in a theory where distances below $l_0$
have no physical meaning. \\
All up to day experiments indicate that $l_0 < 10^{-17} \, cm $, which means that minimal BHs created in a Planckian 
collision, will be certainly quantum objects. Thus, neither classical nor semi-classical description are satisfactory
and one should quantize BHs themselves.\\
In the absence of a proper  quantum mechanical description of BHs, we propose a quantization scheme based on the analogy 
with the quantum harmonic oscillator. This quantization scheme is discussed in Section[\ref{bohr}], where
we also provide a new formulation of QHC. Finally,
in Section[\ref{final}] we summarize the main results obtained.

\section{Self-Regular Schwarzschild solution}
\label{neutral}

In this section we  construct regular Schwarzschild solution of the Einstein equations, where  
the minimal length is dynamically induced, in a self-consistent way. \\
We are looking for a static, spherically symmetric, asymptotically flat metric of the form

\begin{equation}
 ds^2= -\left(\, 1 -\frac{2m(r)}{r}\,\right) dt^2 +\left(\, 1 -\frac{2m(r)}{r}\,\right)^{-1} dr^2 +r^2
\left(\, d\theta^2 + \sin^2\theta d\phi^2\,\right)
\end{equation}

where, $\left(\, r\ ,\theta\ ,\phi\,\right)$ are standard polar coordinates and $t$ is the time measured by
an asymptotic Minkowskian observer.  $m(r)$ is an unknown function determined by the Einstein equations once
the source is given. An energy-momentum tensor compatible with the symmetry of the problem is the one of
an anisotropic fluid:

\begin{eqnarray}
&& T_\mu^\nu = p_\theta \delta_\mu^\nu + \left(\, \rho + p_\theta\,\right) \left(\, u_\mu u^\nu - l_\mu l^\nu\,\right)
\ ,\\
&& p_r +\rho=0\ ,\\
&& T_\mu^\nu{}_{;\nu} =0 \label{divfree}
\end{eqnarray}
In the chosen coordinate system $u^\mu = \delta^\mu_0$, $l^\mu = \sqrt{g_{rr}} \delta^\mu_r$. $\rho$ is the 
energy density, $p_r$ is the radial pressure 
 and $p_\theta$ is the tangential pressure determined in terms of  $\rho$ by the covariant 
divergence-free condition  (\ref{divfree}). \\
From the Einstein equations one finds

\begin{eqnarray}
 m(r)&&= -4\pi \int_0^r dr^\prime r^{\prime\, 2} \,T^{r^\prime}_{r^\prime} \ ,\nonumber\\
      &&=4\pi \int_0^r dr^\prime r^{\prime\, 2} \,\rho(r^\prime)
\label{mradial}
\end{eqnarray}

The textbook Schwarzschild solution for a BH  of mass M is obtained by the choice

\begin{equation}
 \rho=M\frac{\delta(r)}{4\pi r^2} \label{delta}
\end{equation}

which describes a \emph{point-like} source. In our case we choose a smeared matter distribution given by
a Gaussian as:
 
\begin{equation}
 \rho(r)\equiv M \sigma\left(\, r\,\right)
=\left(\frac{3}{l_0}\right)^3 \frac{M}{(4\pi )^{3/2} }\exp\left( -\frac{9 r^2}{4l_0^2}\,\right)
\label{ro}
\end{equation} 

where, $M$ is the total mass-energy  of the system as measured by an asymptotic Minkowskian observer:

\begin{equation}
 M\equiv \lim_{r\to \infty}m(r)= 4\pi \int_0^\infty dr^\prime r^{\prime\, 2} \,\rho(r^\prime)
\end{equation}

This choice draws its motivation from the fact that in ordinary Quantum Mechanics the minimal uncertainty states, 
i.e. the closest states to a point-like object, are given by Gaussian wave-packets.  In the limit
$l_0\to 0$ the function (\ref{ro}) goes into the singular density (\ref{delta}). \\

By inserting (\ref{ro}) in (\ref{mradial}) we obtain

\begin{eqnarray}
 ds^2  &&= -\left(\, 1 -\frac{4M}{{\sqrt\pi} r}\gamma\left(\, 3/2\ , 9r^2/4l_0^2,\right)\,\right) dt^2 +
\left(\, 1 -\frac{4M}{\sqrt{\pi}r}\gamma\left(\, 3/2\ , 9r^2/4l_0^2\,\right)\,\right)^{-1} dr^2 + r^2 d\Omega^2
\ ,\nonumber\\
&&\label{nostra}
\end{eqnarray}

where the incomplete gamma function $\gamma$ is defined as

\begin{equation}
\gamma\left(\, a/b\ ;x\,\right)\equiv
\int_0^x \frac{du}{u}\, u^{a/b} \, e^{-u}
\end{equation}

As a consitency check, we showed the relation $M=M_{ADM}$, where $M_{ADM}$ is the Arnowitt, Deser, Misner mass 
\cite{Arnowitt:1959ah} derived
from the metric (\ref{nostra}). The calculation is straightforward and will be reproduced here.\\  
Horizons correspond to the solutions of the equation

\begin{equation}
g_{rr}^{-1}=0 \longrightarrow M=\frac{\sqrt{\pi}r_h}{4\gamma\left(\, 3/2\ ,9r^2_h /4l_0^2\,\right)}\label{plot}
   \end{equation}
The equation (\ref{plot}) cannot be solved analytically as in the standard Schwarzschild case, but by
plotting the function $M=M(r)$ one sees  the existence of a pair of horizons, merging into a single, degenerate
horizon at the minimum with estimated radius

\begin{figure}[ht!]
\begin{center}
\includegraphics[width=10cm,angle=0]{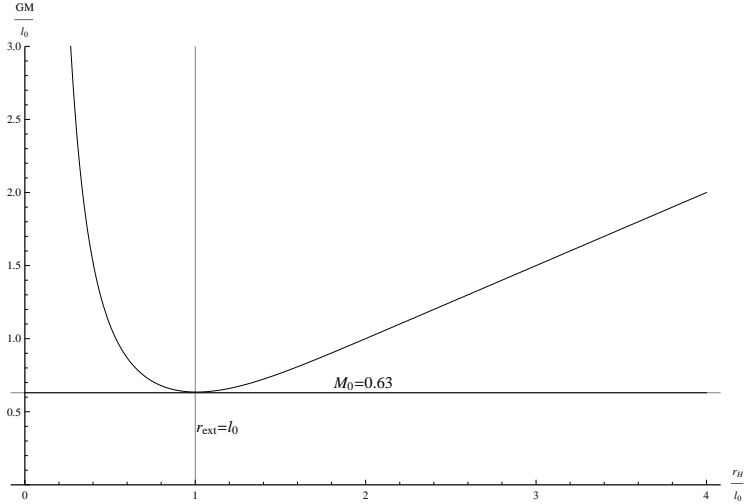}
\caption{\label{qmass} Plot of the mass $M$ as a function of the radius of the horizon. }
\end{center}
\end{figure}

\begin{equation}
 r_{min.}= l_0 +0.01\times l_0
\end{equation}

Neglecting the one per cent corrections, the minimum mass results to be

\begin{equation}
M_0 =\frac{l_0\sqrt{\pi}}{4\gamma\left(\, 3/2\ ,9 /4\,\right)}
\label{m0}
\end{equation}

Thus, for any $M> M_0$ the solution describes a \emph{non-extremal} BH of radius $r_+> l_0$. For $M= M_0$ we have
a \emph{minimal-size, extremal} BH of radius $l_0$ which
gives a physical meaning to the, up to now arbitrarily introduced, cut-off $l_0$. 
 In other words, the existence of a minimal length is a strict consequence of the existence
of minimal size BH of the same radius. This  goes under the name of \emph{self-regular} BH meaning that 
the non-perturbative dynamics of gravity determines a natural
cut-off, thus realizing the UV self-completeness hypothesis in this model.

\section{Bohr quantization of micro BHs}
\label{bohr}
In this section we  present a Bohr-like quantization of BHs. For the sake of simplicity we limit
ourselves to neutral objects only. \\
From the discussion regarding neutral BHs in the Section (\ref{neutral}) 
it results that quantum effects are dominant in
the near extremal region where the behavior of the function $M(r_+)$ significantly differs from the usual 
Schwarzschild case. \\
We shall follow a 
Bohr-like quantization scheme which does not require the knowledge of a full quantum gravity theory. The idea comes 
from the form of (\ref{ro}) which is reminiscent of the  ground-state for an
isotropic, 3D, harmonic oscillator \footnote{A similar idea
has been recently proposed in \cite{Casadio:2013hja} in the framework
of Bose-Einstein condensates model of quantum BHs 
\cite{Dvali:2012gb,Dvali:2011aa,Dvali:2012rt,Dvali:2013vxa,Dvali:2012en} .}

\begin{equation}
\sigma(r)\longleftrightarrow \vert\, \psi_{000}(r)\, \vert^2  
\end{equation} 

where, 

\begin{equation}
 \psi_{000}(r) \propto e^{-m r^2\omega/2} 
\end{equation} 

is the ground state wave function \cite{Spallucci:2014kua}. 
To relate the two different systems, i.e. our BH and the quantum harmonic oscillator, we establish a formal
correspondence between
the mass $m$ and the angular frequency $\omega$ with the corresponding quantities in (\ref{ro}) 

\begin{equation}
 m\omega =\frac{9}{2l_0^2}
\label{omega}
\end{equation} 

In this identification the mass of the extremal BH represents the equivalent of the ground-state energy of the  
harmonic oscillator, i.e. $M_0$ is the zero-point energy

\begin{equation}
 \frac{3}{2}\omega =M_0\label{emme}
\end{equation}

By solving the two equations (\ref{omega}), (\ref{emme}) we find

\begin{equation}
m= \frac{27}{4}\frac{1}{l_0^2 M_0}\ ,\qquad \omega= \frac{2}{3}M_0
\end{equation}
The needs of BH mass quantization has been also recently stressed in \cite{Dvali:2011nh}. Thus, 
we describe non-extremal BHs as ''excited`` energy states labeled  by an integer \emph{principal quantum number} 
$n$ as

\begin{equation}
 M_n =\frac{2}{3} M_0 \left(\, n +\frac{3}{2}\,\right) \ ,\quad n=0\ , 2\ , 4\ ,\dots
\label{hspectrum}
\end{equation} 

Due to the spherical symmetry only even oscillator states are allowed for the 3D, isotropic, harmonic oscillator.
In this quantization scheme
the extremal BH configuration with $r_0=l_0$ represents the \emph{zero-point energy} of the gravitational system.
The result can be interpreted as the realization of earlier attempts  to dynamically generate a
\emph{zero-point length} \cite{Padmanabhan:1996ap,Fontanini:2005ik,Spallucci:2005bm,Kothawala:2014fva} 
thus eliminating  ultraviolet divergences through quantum fluctuations of gravity itself. 
 \\
Following further  analogy with Bohr quantization, where the quantum/classical transition is achieved for 
large $n$, we redefine a ''quantum`` mass/energy distribution which for large $n$ approaches a  Dirac delta 
sourcing a standard Schwarzschild metric \cite{Nicolini:2005vd,Rizzo:2006zb,Nicolini:2008aj}.  
In other words, we are adopting a kind of Correspondence
Principle, \emph{a la} Bohr, applied to the matter energy density:

\begin{equation}
 \rho_n(r)\equiv M_n \sigma_n\left(\, r\,\right)
= \left(\frac{n+3/2}{l_0}\right)^3 \frac{M_n}{\pi^{3/2} }
\exp\left[ -\frac{r^2}{l_0^2}\left(\, n + 3/2\,\right)^2\,\right]
\label{ron}
\end{equation} 

Notice that

\begin{equation}
 \rho_n(r)\longrightarrow M_n \, \frac{\delta(r)}{4\pi r^2}\ , \quad n>> 1
\end{equation}

leading to a standard Schwazschild geometry. One may wonder why we do not push the analogy to an extreme and use 
the known excited state wave-functions of the harmonic oscillator. We considered this approach but did not pursue it
for the following reasons:
\begin{itemize}
 \item  we called our BH quantization conjecture ''Bohr-like``, in the same spirit in which old quantum mechanics
was formulated  before Schroedinger, since there is no wave-equation for quantum BHs as there is for quantum harmonic
oscillator.
\item  Even if one ignores the previous comment, the use of excited harmonic oscillator wave-functions leads to
multi-horizon geometries with an increasing number of different extremal configurations. The resulting
excited BHs have a geometrical structure completely different from
the minimal size, self-regular,  solution we started from. 
\end{itemize}

Therefore, the ``quantized'' version  of (\ref{nostra}) reads

\begin{figure}[ht!]
\begin{center}
\includegraphics[width=10cm,angle=0]{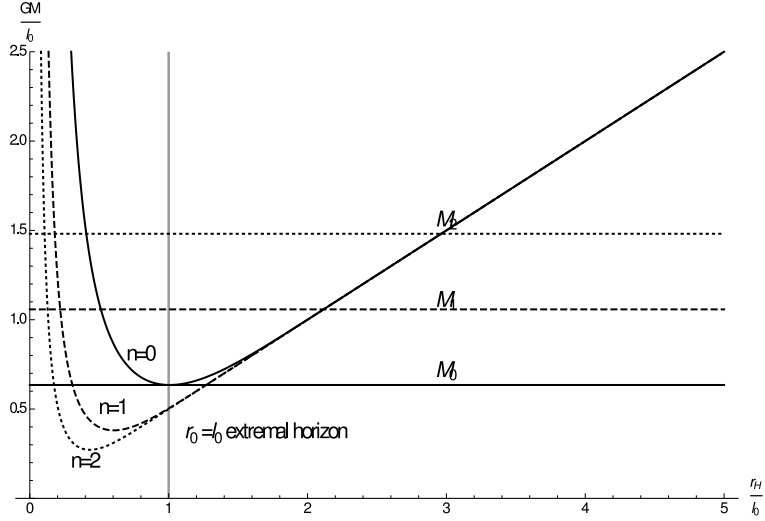}
\caption{Plot of the function $\mathcal{M}_n(r)$ for  $n=0\ , 2\ , 4$. Horizontal lines correspond to quantized mass 
levels. }
\end{center}
\end{figure}

 \begin{eqnarray}
 && ds^2 = -g_{00}\, dt^2 + g_{00}^{-1}\, dr^2 + r^2 d\Omega^2\label{dsn}\\
 && g_{00}  = 1-\frac{2\mathcal{M}_n\left(\, r\,\right)}{r} \ ,\\
 && \mathcal{M}_n\left(\, r\,\right) =M_n
\gamma\left[\, 3/2\ , \frac{r^2}{l_0^2}\left(\, n+3/2\,\right)^2\,\right]/\Gamma(3/2)
\label{qmetric}
\end{eqnarray} 

Several comments are in order.\\
First, due to the identification of the extremal BH with the $n=0$ state, we obtain the
\emph{zero-point metric},  with quantization picking up only the ground-state mass $M=M_0$.\\
Second,  the metric (\ref{qmetric}), for ``~large-$n$~'' reduces to the ordinary Schwarzschild  
line element while still keeping a quantized mass spectrum.

\begin{equation}
 ds^2 =-\left(\, 1 -\frac{2M_n}{r} \,\right) dt^2 
+\left(\, 1 -\frac{2M_n}{r} \,\right)^{-1}dr^2 + r^2 d\Omega^2
\label{large}
\end{equation}

where, $M_n \approx 2n M_0/3 $.  This limit is due to the fact $\gamma ( \, n \; x\, )\longrightarrow \Gamma(n)$ for
$n$ large enough. As a matter of fact, already for $n>2$ the metric (\ref{large}) is  a very good approximation of 
the exact quantum metric (\ref{qmetric}). The relative energy difference between nearby levels  
$\Delta M_n/M \sim 1/n$ for $n>>1$ and the mass spectrum becomes \emph{effectively} continuous.\\

Two limiting cases are of particular interest: ground state $n=0$, and the
\emph{classical (large BH)} limit  $n>> 1$. \\
On expects that the ground state of the system is only ``vacuum energy'', i.e. that the extremal BH configuration
is only a vacuum-fluctuation. 
From the effective geometry (\ref{qmetric}) one obtains the ``semi-classical'' horizon equation 
\begin{equation}
 r_+ =2M_n\,\gamma\left(\, 3/2\ ,  \frac{r^2_+}{l_0^2}
\left(\, n+ 3/2\,\right)^2 \,\right) / \Gamma(3/2)
\label{reff}
\end{equation}

which we translate into a quantum framework as the equation for the average values of horizon radius in a given
quantum state.  We identify the trivial solution $r_+=0$ as the vacuum average value:

\begin{equation}
 <\,0\,\vert r_+\,\vert \,0\,> =0 \Leftrightarrow r_+=0
\label{zero}
\end{equation}

However, this vanishing mean value has an ``uncertainty'' :

\begin{equation}
 \Delta r_+ =\sqrt{< \, 0\, \vert\, r_+^2\, \vert\, 0 > }
\end{equation}

By squaring (\ref{reff}) we get the equation for the
vacuum average of $r_+^2$:

\begin{equation}
< \, 0\, \vert\, r_+^2\, \vert\, 0\, > =\left[\, 2M_0\,
\gamma\left(\, 3/2\ ,  \frac{9<\, 0\,\vert \,r^2_+\,\vert\, 0\, >}{4l_0^2}
 \,\right) / \Gamma(3/2)\,\right]^2
\label{fluct}
\end{equation}

By taking into account the definition (\ref{m0})  of $M_0$ it is immediate to check that equation (\ref{fluct}) is
solved by

\begin{equation}
 < \, 0\, \vert\, r_+^2\, \vert\, 0 >= l_0^2
\label{v2}
\end{equation}

and $ \Delta r_+ =l_0 $.\\
In the large-$n$ limit equation (\ref{reff}) reduces to the definition of the classical Schwarzschild radius

\begin{equation}
 <\,n \vert \, r_+\,\vert n>=2M_n\ , \qquad n>> 1
\label{rs}
\end{equation}
and
\begin{equation}
 <\,n \vert \, r_+^2\,\vert n>=4M_n^2 \longrightarrow \Delta r_+ \to 0
\label{rs2}
\end{equation}

As it was expected, the extremal BH, corresponding to the zero-point energy of the system, is a pure
quantum fluctuation, while highly excited states behave as ``classical'' objects ($\Delta r_+ =0$)
described by an effective Schwarzschild metric.
Thanks to the properties of the $\gamma$-function,  a good approximation of the full spectrum is

\begin{equation}
 <\,n \vert \, r_+\,\vert n>\simeq 2M_n\,\gamma\left(\, 3/2\ , 
n^2 \,\right) / \Gamma(3/2)
\label{spettro}
\end{equation}

\subsection{Quantum hoop conjecture}
Strictly speaking, the density (\ref{ron}) is non-vanishing everywhere, even if it quickly drops to zero
already at distances of few $l_0$. Nevertheless,   the skeptics may rise the question whether a  BH can be 
formed at all by such a smeared distribution. 
In order to remove these doubts, we evoke the classical \emph{hoop conjecture} \cite{thorne} and adapt it
to the present situation \footnote{A quantum formulation of the hoop conjecture has been recently proposed in 
\cite{Casadio:2013uga}.}. 
First, we define a \emph{mean radius} of the mass distribution and the mean value of the square radius, as

\begin{eqnarray}
&& <n\,\vert\, r\,\vert \, n > \equiv 4\pi \int_0^\infty dr r^3 \sigma_n(r) 
                      = \frac{l_0}{(n+3/2)}\frac{1}{\Gamma(3/2)}\ ,\\
&& <n\,\vert\, r^2\,\vert \, n >\equiv 4\pi \int_0^\infty dr r^4 \sigma_n(r)
                  = \frac{l_0^2}{\left(\, n+3/2\,\right)^2}\frac{\Gamma\left(\,\frac{5}{2}\,\right)}{\Gamma(3/2)}
\end{eqnarray}

Secondly, we evaluate the mean square deviation as

\begin{equation}
 \Delta r_n^2\equiv  <n\,\vert\, r^2\,\vert \, n >- <n\,\vert\, r\,\vert \, n >^2= 
\frac{l_0^2}{\left(\, n+3/2\,\right)^2}\left[\,\frac{3\pi}{8}-1\,\right]
\end{equation}

which vanishes for large $n$: $\Delta r_n \longrightarrow 0$.\\

Finally, we define the quantum hoop conjecture as the condition that \emph{whenever the mean radius of the mass 
distribution is smaller that the mean value of the horizon radius, then a quantum BH forms}. 

\begin{equation}
 < n\,\vert \, r\,\vert\, n>\, \le \,< n\,\vert \, r_+\,\vert\, n>
\end{equation}  

which leads to the relation

\begin{equation}
 \frac{l_0}{n+ 3/2}\le 2M_n \gamma\left[\, 3/2\ ; n^2\,\right] 
\label{qhoop}
\end{equation}

It is sufficient to verify (\ref{qhoop}) in the ``worst case scenario'' $n=1$, which turns out to be
satisfied. For larger $n$
the width of the Gaussian distribution shrinks while the radius of the horizon increases, maintaining the QHC. 

\section{Conclusions and discussion}
\label{final}
In the first part of this letter  we described a regular Schwarzschild geometry, 
incorporating a  ``minimal'' length $l_0$ which, due to gravitational quantum dynamics, 
turns out to be the radius of the minimal size extremal 
configuration. This solution is a first realization, within a specific model, of the UV self-complete quantum gravity 
program. In other words, quantum gravity effects dynamically generate a short distance cut-off shielding the Planck 
scale physics from experimental probe. 
In the last part of this work, we have proposed a Bohr-like BH quantization scheme. For the sake of computational
simplicity, we have described  neutral quantum BHs. The generalization to charged BHs will follow the same pattern
in a bit more technically involved manner.\\ 
   The main outcome of the proposed quantization scheme can be listed as:
\begin{itemize}
 \item  any phenomenon occurring at distances smaller than
$l_0$ is not experimentally measurable.
 \item The mass of the extremal configuration is equivalent to the 
       zero-point energy of the quantum harmonic oscillator.
 \item The non-extremal configurations corresponds to harmonic oscillators excited states, with a discrete mass 
       spectrum of equally spaced levels.
\item  The model satisfies a Bohr-like Correspondence Principle for large $n$, where it reproduces standard 
Schwarzschild BH.  
\item  We also formulated a ``quantum hoop conjecture'', which  supports the existence of quantum BHs whenever 
the condition 
$  <n\,\vert\, r\,\vert \, n >\le  <n\,\vert\, r_+\,\vert \, n > $ is met.
\item One may wonder what is the thermodynamics of quantum BHs. We believe that concepts like Hawking temperature 
and Bekenstein entropy refer to semi-classical BHs where mass and size are continuous variables. Intrinsically
quantum BHs do not radiate thermally being  stationary state configurations. Non-thermal quantum BHs have been
recently discussed in \cite{Calmet:2008dg}. In the quantum phase absorption and emission proceed through
discrete quantum jumps between different energy states, instead of emitting a continuous thermal spectrum.  Thus,
there is no Hawking radiation at the quantum level and BHs are just another kind of  ``particle'' in the quantum zoo.
\end{itemize}


\begin{thebibliography}{99}

\bibitem{Delbourgo:1969ha} 
  R.~Delbourgo, A.~Salam and J.~A.~Strathdee,
  Lett.\ Nuovo Cim.\  {\bf 2}, 354 (1969)
\bibitem{Salam:1970bd} 
  A.~Salam and J.~A.~Strathdee,
 Lett.\ Nuovo Cim.\  {\bf 4}, 101 (1970)
\bibitem{Isham:1970aw} 
  C.~J.~Isham, A.~Salam and J.~A.~Strathdee,
  Phys.\ Rev.\ D {\bf 3}, 1805 (1971) 

\bibitem{SriRam:1979fr} 
  M.~S.~Sri Ram and T.~Dass,
  Phys.\ Rev.\ D {\bf 22}, 814 (1980).


\bibitem{Dvali:2010bf} 
  G.~Dvali and C.~Gomez,
  ``Self-Completeness of Einstein Gravity,''
  arXiv:1005.3497 [hep-th].


\bibitem{Dvali:2010jz} 
  G.~Dvali, G.~F.~Giudice, C.~Gomez and A.~Kehagias,
  JHEP {\bf 1108}, 108 (2011)

\bibitem{thorne}
  K.~S.~Thorne, {\em Nonspherical gravitational collapse, a short review},
in *J R Klauder, Magic Without Magic*, Freeman, San Francisco 1972, 231-258.

\bibitem{Casadio:2013uga} 
  R.~Casadio, O.~Micu and F.~Scardigli,
  Phys.\ Lett.\ B {\bf 732}, 105 (2014)

\bibitem{Rizzo:2002kb} 
  T.~G.~Rizzo,
  JHEP {\bf 0202}, 011 (2002)

\bibitem{Chamblin:2002ad} 
  A.~Chamblin and G.~C.~Nayak,
  Phys.\ Rev.\ D {\bf 66}, 091901 (2002)

\bibitem{Mocioiu:2003gi} 
  I.~Mocioiu, Y.~Nara and I.~Sarcevic,
  Phys.\ Lett.\ B {\bf 557}, 87 (2003)

\bibitem{Lonnblad:2005ah} 
  L.~Lonnblad, M.~Sjodahl and T.~Akesson,
  JHEP {\bf 0509}, 019 (2005)

\bibitem{Rizzo:2006di} 
  T.~G.~Rizzo,
  Phys.\ Lett.\ B {\bf 647}, 43 (2007)

\bibitem{Najafabadi:2008xv} 
  M.~Mohammadi Najafabadi and S.~Paktinat Mehdiabadi,
  JHEP {\bf 0807}, 011 (2008)

\bibitem{Chamblin:2008ec} 
  A.~Chamblin, F.~Cooper and G.~C.~Nayak,
  Phys.\ Lett.\ B {\bf 672}, 147 (2009)

\bibitem{Erkoca:2009kg} 
  A.~E.~Erkoca, G.~C.~Nayak and I.~Sarcevic,
  Phys.\ Rev.\ D {\bf 79}, 094011 (2009)

\bibitem{Kiritsis:2011qv} 
  E.~Kiritsis and A.~Taliotis,
  ``Mini-Black-Hole Production at RHIC and LHC,''
  PoS EPS {\bf -HEP2011}, 121 (2011)

 

\bibitem{Spallucci:2011rn} 
  E.~Spallucci and S.~Ansoldi,
  Phys.\ Lett.\ B {\bf 701}, 471 (2011)

\bibitem{Mureika:2011hg} 
  J.~Mureika, P.~Nicolini and E.~Spallucci,
  Phys.\ Rev.\ D {\bf 85}, 106007 (2012)

\bibitem{Spallucci:2012xi} 
  E.~Spallucci and A.~Smailagic,
  Phys.\ Lett.\ B {\bf 709}, 266 (2012)

\bibitem{Nicolini:2011fy} 
  P.~Nicolini, A.~Orlandi and E.~Spallucci,
  Adv.\ High Energy Phys.\  {\bf 2013}, 812084 (2013)


\bibitem{Aurilia:2013mca} 
  A.~Aurilia and E.~Spallucci,
  Adv.\ High Energy Phys.\  {\bf 2013}, 531696 (2013)

\bibitem{Nicolini:2012fy} 
  P.~Nicolini and E.~Spallucci,
  Adv.\ High Energy Phys.\  {\bf 2014}, 805684 (2014)

\bibitem{Nicolini:2013ega} 
  P.~Nicolini, J.~Mureika, E.~Spallucci, E.~Winstanley and M.~Bleicher,
  ``Production and evaporation of Planck scale black holes at the LHC,''
  arXiv:1302.2640 [hep-th].

\bibitem{Dvali:2011th} 
  G.~Dvali, C.~Gomez and A.~Kehagias,
  JHEP {\bf 1111}, 070 (2011)

\bibitem{Dvali:2012mx} 
  G.~Dvali and C.~Gomez,
  JCAP {\bf 1207}, 015 (2012)
  
\bibitem{Dvali:2014ila} 
  G.~Dvali, C.~Gomez, R.~S.~Isermann, D.~Lust and S.~Stieberger,
  ``Black Hole Formation and Classicalization in Ultra-Planckian $2 \to N$ Scattering,''
  arXiv:1409.7405 [hep-th].

\bibitem{book1}
A.~Hagar,
``~Discrete or Continuous?: The Quest for Fundamental Length in Modern Physics~''
Cambridge University Press ( 2014)

\bibitem{Moffat:2000fv} 
  J.~W.~Moffat,
  Phys.\ Lett.\ B {\bf 493}, 142 (2000)

\bibitem{Moffat:2000gr} 
  J.~W.~Moffat,
  Phys.\ Lett.\ B {\bf 491}, 345 (2000)



\bibitem{Nicolini:2005vd} 
  P.~Nicolini, A.~Smailagic and E.~Spallucci,
  Phys.\ Lett.\ B {\bf 632}, 547 (2006)

\bibitem{Rizzo:2006zb} 
  T.~G.~Rizzo,
  JHEP {\bf 0609}, 021 (2006)

\bibitem{Ansoldi:2006vg} 
  S.~Ansoldi, P.~Nicolini, A.~Smailagic and E.~Spallucci,
  Phys.\ Lett.\ B {\bf 645}, 261 (2007)


\bibitem{Nicolini:2008aj} 
  P.~Nicolini,
  Int.\ J.\ Mod.\ Phys.\ A {\bf 24}, 1229 (2009)

\bibitem{Spallucci:2009zz} 
  E.~Spallucci, A.~Smailagic and P.~Nicolini,
  Phys.\ Lett.\ B {\bf 670}, 449 (2009)

\bibitem{Smailagic:2010nv} 
  A.~Smailagic and E.~Spallucci,
  Phys.\ Lett.\ B {\bf 688}, 82 (2010)


\bibitem{Arnowitt:1959ah} 
  R.~L.~Arnowitt, S.~Deser and C.~W.~Misner,
  Phys.\ Rev.\  {\bf 116}, 1322 (1959).




\bibitem{Dvali:2011nh} 
  G.~Dvali, C.~Gomez and S.~Mukhanov,
  ``Black Hole Masses are Quantized,''
  arXiv:1106.5894 [hep-ph].


\bibitem{Padmanabhan:1996ap} 
  T.~Padmanabhan,
  Phys.\ Rev.\ Lett.\  {\bf 78}, 1854 (1997)

\bibitem{Fontanini:2005ik} 
  M.~Fontanini, E.~Spallucci and T.~Padmanabhan,
  Phys.\ Lett.\ B {\bf 633}, 627 (2006)

\bibitem{Spallucci:2005bm} 
  E.~Spallucci and M.~Fontanini,
  ``Zero-point length, extra-dimensions and string T-duality,''
  gr-qc/0508076.

\bibitem{Kothawala:2014fva} 
  D.~Kothawala and T.~Padmanabhan,
  arXiv:1408.3963 [gr-qc].

\bibitem{Casadio:2013hja} 
  R.~Casadio and A.~Orlandi,
  JHEP {\bf 1308}, 025 (2013)

\bibitem{Dvali:2012gb} 
  G.~Dvali and C.~Gomez,
  Phys.\ Lett.\ B {\bf 716}, 240 (2012)

\bibitem{Dvali:2011aa} 
  G.~Dvali and C.~Gomez,
  Fortsch.\ Phys.\  {\bf 61}, 742 (2013)



\bibitem{Dvali:2012rt} 
  G.~Dvali and C.~Gomez,
  Phys.\ Lett.\ B {\bf 719}, 419 (2013)

\bibitem{Dvali:2013vxa} 
  G.~Dvali, D.~Flassig, C.~Gomez, A.~Pritzel and N.~Wintergerst,
  Phys.\ Rev.\ D {\bf 88}, no. 12, 124041 (2013)

\bibitem{Dvali:2012en} 
  G.~Dvali and C.~Gomez,
  Eur.\ Phys.\ J.\ C {\bf 74}, 2752 (2014)

\bibitem{Spallucci:2014kua} 
  E.~Spallucci and A.~Smailagic,
  ``Semi-classical approach to quantum black holes'' in
 "Advances in black holes research",p.1-25, Ed. A.Barton, Nova Science Publishers, Inc., (2015);
 arXiv:1410.1706 [gr-qc]
\bibitem{Calmet:2008dg} 
  X.~Calmet, W.~Gong and S.~D.~H.~Hsu,
  Phys.\ Lett.\ B {\bf 668}, 20 (2008)

\end{thebibliography}
\end{document}